# Computer Folding of RNA Tetraloops: Identification of Key Force Field Deficiencies

*Petra Kührová,[§] Robert B. Best,[‡] Sandro Bottaro,[†] Giovanni Bussi,[†] Jiří Šponer,[§£]\* Michal Otyepka[§£] and Pavel Banáš,[§£]\**

[§]Regional Centre of Advanced Technologies and Materials, Department of Physical Chemistry, Faculty of Science, Palacky University Olomouc, 17. listopadu 12, 771 46 Olomouc, Czech Republic e-mail: pavel.banas@upol.cz

[‡]Laboratory of Chemical Physics, National Institute of Diabetes and Digestive and Kidney Diseases, National Institutes of Health, Bethesda, MD 20892-0520

[†]Scuola Internazionale Superiore di Studi Avanzati, Via Bonomea 265, 34136 Trieste, Italy

[£]Institute of Biophysics, Academy of Sciences of the Czech Republic, Kralovopolska 135, 612 65 Brno, Czech Republic

*corresponding authors

KEYWORDS: force field, RNA tetraloop, folding, enhanced sampling methods, REMD

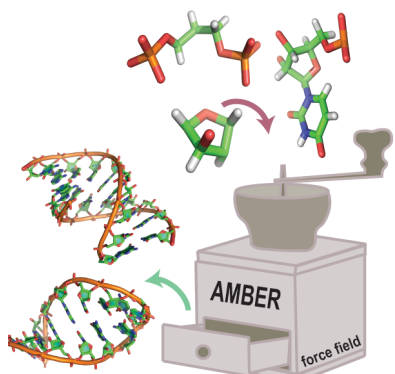

Table of Contents: The developments of enhanced sampling methods and force fields are highly mutually interrelated efforts. Here we used three different enhanced sampling techniques (temperature based replica exchange molecular dynamics, replica exchange solute tempering, and metadynamics) to fold 5'-GAGA-3' RNA tetraloop. We aimed to separate problems caused by limited sampling from those due to force-field inaccuracies, and identify which terms of the force field are responsible for poor description of tetraloop folding.

**ABSTRACT**

The computer aided folding of biomolecules, particularly RNAs, is one of the most difficult challenges in computational structural biology. RNA tetraloops are fundamental RNA motifs playing key roles in RNA folding and RNA-RNA and RNA-protein interactions. Although state-of-the-art Molecular Dynamics (MD) force fields correctly describe the native state of these tetraloops as a stable free-energy basin on the microsecond time scale, enhanced sampling techniques reveal that the native state is not the global free energy minimum, suggesting yet unidentified significant imbalances in the force fields. Here we tested our ability to fold the RNA tetraloops in various force fields and simulation settings. We employed three different enhanced sampling techniques, namely temperature replica exchange MD (T-REMD), replica exchange with solute tempering (REST2), and well-tempered metadynamics (WT-MetaD). We aimed to separate problems caused by limited sampling from those due to force-field inaccuracies. We found that none of the contemporary force fields is able to correctly describe folding of the 5'-GAGA-3' tetraloop over a range of simulation conditions. We thus aimed to identify which terms of the force field are responsible for this poor description of TL folding. We showed that at least two different imbalances contribute to this behavior, namely, overstabilization of base-phosphate and/or sugar-phosphate interactions and underestimated stability of the hydrogen bonding interaction in base pairing. The first artifact stabilizes the unfolded ensemble while the second one destabilizes the folded state. The former problem might be partially alleviated by reparameterization of the Van der Waals parameters of the phosphate oxygens suggested by Case et al., while in order to overcome the latter effect we suggest local potentials to better capture hydrogen bonding interactions.

# INTRODUCTION

Computer simulation of the folding of the biomolecules is one of the major challenges in biomolecular modelling. Simulations, together with experimental methods, help us to build a more complete picture of these nontrivial processes.[1-7] Computational studies of folding often use molecular dynamics (MD) simulations in conjunction with enhanced sampling methods. The accuracy, and in turn the predictive power of folding simulations critically depends on the quality of enhanced sampling techniques, accessible time scales, and most importantly on the force fields used. Although modern force fields often perform satisfactorily in unbiased simulations initiated from RNA experimental structures, this does not guarantee that the native state is the global free energy minimum within the force field description, which would be required to achieve folding from the unfolded state. Therefore, contemporary force fields should be validated not only using unbiased MD simulations, which are typically trapped in a free energy basin corresponding to the starting structure, but also using enhanced sampling methods capable to describe e.g. folding of the RNA. The developments of enhanced sampling methods and force fields are thus highly mutually interrelated efforts. The enhanced sampling methods are designed to overcome energy barriers and provide a robust exploration of free energy landscape.[8] In the limit of converged sampling, the accuracy of enhanced sampling simulations is determined by the accuracy of the force field.[9]

There are many variants of enhanced sampling MD methods. The widely used temperature-based replica exchange MD[10] (T-REMD) benefits from multiple independent MD simulations running in parallel over a range of different temperatures. Exchanges of configurations between neighbouring temperatures are attempted at fixed time intervals and accepted according to a Metropolis-style algorithm to ensure canonical sampling at all temperatures. Another method based on multiple independent MD simulations is replica exchange with solute scaling (in a REST2 variant, see ref. [11]), which is a variant of Hamiltonian-based replica exchange MD (H-REMD).[12] This method is based on a modification of the potential energy, so that the interactions between solute atoms are scaled by a factor $\lambda$, solvent-solvent interactions remain unscaled, and solute-solvent interactions are scaled by an intermediate factor (in this case by $\sqrt{\lambda}$). Scaling the energy by a factor $\lambda$ is equivalent to a scaling of the temperature by $1/\lambda$. Thus, in case of REST2 only the solute atoms are effectively heated up. What is more important, the solvent-solvent interactions that typically contribute the most to the energy differences between replicas in T-REMD simulations do not contribute to exchanges. Another popular and conceptually very different enhanced sampling method is a well-tempered metadynamics (WT-MetaD).[13-14] WT-MetaD uses a history-dependent biasing potential on a few chosen coarse-grained degrees of freedom referred to collective variables (CVs). Suitable CVs help the systems to escape from the trap of the free energy minima and accelerate sampling of rare events. The well-tempered version of the metadynamics prevents overfilling of the free energy landscape and generally improves convergence of this technique.[13] Defining an appropriate set of CVs, which would allow the most relevant conformational changes to be described, is usually the most difficult challenge.[15-16]

RNA hairpins composed of a short loop at the tip of the Watson-Crick base-paired A-RNA stem are ideal testing systems for validation of force fields because of their small size and the significance of interactions other than Watson-Crick base pairs in stabilizing their structure. These abundant building blocks are indispensable for RNA folding[17-18] and play crucial roles in RNA-RNA and protein-RNA interactions. Hairpin loops containing four nucleotides in the loop region, denoted tetraloops (TLs), are notably stable and are often involved in tertiary contacts.[17, 19-22] Moreover, TLs play several biological roles in translation and transcription.[17] They have even been suggested to be involved in the emergence of spontaneous RNA catalysis in early stages of RNA prebiotic chemistry.[23] The most prominent families of TLs are 5'-GNRA-3' and 5'-UNCG-3' TLs (N stands for any nucleotide and R for purine). GNRA TLs can participate in protein-RNA complexes serving as a recognition sites for RNA binding and are involved in many RNA-RNA and RNA-ligand interactions.[24]

Experimental results indicate that the RNA TL folding is more hierarchical compared to folding of small fast-folding proteins.[25-26] Two types of RNA TL folding mechanism have been suggested.[27] In the first model, the RNA chain collapses through non-specific base pairs to form a heterogeneous structural ensemble, followed by native-like or misfolded helix growth from the nucleation centre.[27-28] The second pathway is a zipping mechanism, in which the loop is closed by stable base pair, followed by subsequent zipping of the helical stem.[27] Despite their small size, a computational description of the folding of RNA TLs remains a challenge.[29] RNA TLs are precisely-shaped molecular building blocks with characteristic signature molecular interactions determining their consensus sequences.[30] The TL folding process results from a complex interplay of canonical base pairing, non-canonical interactions, stacking, solvation effects, backbone substates and electrostatic interactions between RNA and surrounding ions. The main obstacles to the accurate molecular description of the TLs folding are limited sampling in molecular simulations and limited accuracy of the force field.[31]

Current atomistic simulations of nucleic acids are still based on second-generation pair-additive force fields derived around twenty years ago.[2-3] There have been efforts to improve their performance by partial reparametrizations.[32-38] Most of these works attempted tuning of the uncoupled one-dimensional backbone dihedral potentials, which is the most straightforward refinement. These rather unphysical tweaks are used for final tuning of the force field once the other parameters are fixed, and we certainly cannot expect that they can provide a "perfect" force field. Rather, they are used to eliminate the most drastic force field failures, such as degradation of canonical helices. Regarding specifically RNA, one such refinement is the CHARMM36 tuning of the 2'-OH group dihedral parameters.[39] However, the CHARMM RNA force field still suffers from some understabilization of canonical A-RNA helices on the 10-100 ns time scale.[40-41] Thus, in contrast to good performance of CHARMM force field in the proteins and B-DNA simulations, variants of the AMBER Cornell et al. force field[42] are usually preferred for long RNA simulations. Due to the known principal limitations of pair-additive force fields, the efforts in CHARMM force field development has recently shifted to the promising design of a polarization force field, though its RNA version is not yet available.[43]

In the case of the AMBER nucleic acid force field, the 2007 bsc0 reparameterization[35] eliminated spurious α/γ flips resulting in a progressive degradation of B-DNA structure during simulations. Although bsc0 is unnecessary for RNA simulations, longer simulations have proved that it also improves the RNA description.[32, 40, 44] In 2010 we identified a similar serious problem in RNA simulations, namely the formation of ladder-like structures in canonical A-form helices.[45] Subsequently we derived a reparameterization of the glycosidic torsion $\chi_{OL3}$[32, 38] which supressed the ladder-like structures and became a part of the contemporary standard AMBER force field. Suppression of the RNA ladder-like structure required the DNA and RNA AMBER force fields to be separated, due to incompatible requirements on the $\chi$ dihedral potentials in DNA and RNA. The bsc0$\chi_{OL3}$ variant was tested on many RNA systems including RNA helices[40, 44] and RNA TLs.[29, 31-32] No undesired side-effects for this specific modification have been found so far, though the force field as a whole of course remains far from "perfect". Another reparameterization of the χ torsion was independently suggested by Yildirim et al., though the authors were not aware of the ladder problem.[36] We have shown that this reparameterization also prevents the spurious ladder-like structures albeit with moderate side effects because the potential in the *anti* to high-*anti* region is too steep. This causes some flattening of the A-form helix by underestimation of inclination and roll parameters.[32] Subsequently, Yildirim et al. extended their version by including bsc0 and reparameterization of the ε, ζ, and β torsions (known as AmberTOR[46]). At least in our tests, this complete dihedral reparametrization caused canonical A-form helices to deteriorate.[47] In 2013, Zgarbova et al. derived a DNA-specific reparameterization of ε/ζ torsions denoted εζ$_{OL1}$ that significantly improved the description of B-DNA and DNA G-quadruplexes.[37] Subsequently we showed that εζ$_{OL1}$ also significantly improved the description of non-canonical sugar-phosphate backbone conformation at the active site of hairpin ribozyme.[34] However, the general applicability of εζ$_{OL1}$ to RNA has not yet been clarified. Additionally, our study also indicated that RNA simulations might be improved by implementation of modified Van der Waals parameters of the phosphate group, as suggested by Case et al.[34, 48] This was later supported by Cheatham et al. in REMD simulations of tetranucleotides.[49] An alternative reparameterization of ff99 AMBER force field has been suggested by Chen&Garcia (further referred as ff99Chen) during their efforts to achieve a reversible folding of RNA tetraloops.[33] It has included modification of van der Waals parameters of bases with additional atom-pair specific modification of the Van der Waals combination rules for water-base interactions using the so-called NBfix scheme. The main aim was to eliminate the known overestimation of the stacking interactions.[50-51] It has been accompanied by an adjustment of the χ dihedral potential. The original T-REMD study reported multiple folding events to the folded state with correct signature interactions for two out of three studied RNA TLs, including a GNRA TL. A subsequent benchmark study by Cheatham et al. (multi-dimensional REMD with restrained A-RNA stem below the loop) confirmed that the ff99Chen force field leads to an improvement over the standard AMBER force field for the TLs. Specifically, the ff99Chen force field had clearly the highest population of the native UUCG loop conformation among all the tested force fields (~10% and ~1% at 277 K and 300 K, respectively), based on analysis of the signature interactions.[52] On the other hand, the ff99Chen force field may lead to excessive stabilization of some non-native base-pairs and flattening (i.e., loss of inclination) of A-RNA in some other systems, underlining the fact how difficult is to obtain a force field that would simultaneously reproduce all properties of RNA molecules.[52-53] Some studies also noted that nucleic acids simulations are somewhat sensitive to the choice of explicit water model,[40, 44, 54] though variations of water models are unlikely to resolve the limited accuracy of the primary solute force field, which originates from its inability to reliably describe inherent conformational preference of nucleic acids.[55] In any case, due to the above-noted limited capability to fine-tune the force fields by dihedral potential reparametrizations, future efforts to modify the AMBER non-bonded terms are expected.

In this work, we study the conformational dynamics and folding of a small RNA hairpin, namely 5'-GAGA-3' TL. Our study investigates the effect of the oligonucleotide length, ionic conditions, choice of the sampling technique, force field, and the definition of the folded native state on the description of the folding free energy landscape. The main aim is to identify possible force field artefacts precluding correct theoretical description of the TL folding. In total, we report eight T-REMD simulations with a cumulative time of 896 μs, one REST2 simulation with a cumulative time of 32 μs, and two 3-μs-long WT-MetaD simulations, i.e., almost 1 millisecond of simulations on aggregate.

**METHODS**
STARTING STRUCTURES
**The folded state.** The folded state of GAGA TL was taken from the 1.04 Å resolution X-ray structure of the sarcin-ricin loop from *E. coli* 23S ribosomal RNA (PDB ID 1Q9A[56], residues 2658-2663) and capped by an additional one or two CG base-pairs yielding the final sequences 5'-gcGAGAgc-3' or 5'-cgcGAGAgcg-3', respectively. The native structure of GAGA TL, shown in Fig. 1, is stabilized by three H-bonds: $G_{L1}(N2)\cdots A_{L4}(pro\text{-}R_p)$ (3BPh interaction;[57] it can alternate with $G_{L1}(N1/N2)\cdots A_{L4}(pro\text{-}R_p)$ bifurcated H-bonds 4BPh interaction during the simulation), $G_{L1}(N2)\cdots A_{L4}(N7)$, and $G_{L1}(O2')\cdots G_{L3}(N7)$. The $A_{L2}$, $G_{L3}$, and $A_{L4}$ bases form a purine triple base stack and $G_{L1}$ is base-paired with $A_{L4}$ by the *trans* Hoogsten/Sugar-Edge (*t*HS) $A_{L4}/G_{L1}$ pattern.

**The unfolded states.** The initial structure of a single strand was built up using the Nucleic Acid Builder package of AmberTools14[58] as one strand of an A-form duplex. Subsequently, a starting ensemble of different replicas was prepared using classical 640-ns-long MD simulation of the single strand of the 8-mer and 10-mer at 300 K, initiated from the above-mentioned structure. The 64 unfolded structures were taken in 10 ns long intervals. The protocol of the classical MD simulation of the unfolded state is described in Supporting Information.

The starting topology and coordinates of folded and unfolded states of GAGA TL were prepared using the tLEaP module of AMBER 12 program package. The structures were solvated using a rectangular box of TIP3P[59] water molecules with a minimum distance between box walls and solute of 10 Å for the unfolded state and 15 Å for the folded state, respectively, yielding ~7000 water molecules added and ~60×60×60 Å³ box sizes. The simulations were performed in either ~150 mM KCl salt or in $K^+$ net-neutral conditions. We used the Joung-Cheatham ionic parameters[60] ($K^+$: $r = 1.705$ Å, $\varepsilon = 0.1937$ kcal/mol, $Cl^-$: $r = 2.513$ Å, $\varepsilon = 0.0356$ kcal/mol) in all simulations except for the T-REMD simulation with the ff99Chen force field,[33] where we used the suggested parameterization of ion-ion and ion-water interactions developed by Chen and Pappu.[61] In WT-MetaD simulations we used $Na^+$ net-neutral condition with Åqvist parameters[62] ($r = 1.87$ Å and $\varepsilon = 0.0028$ kcal/mol). All structures were minimized and equilibrated using the standard equilibration protocol described in Supporting Information.

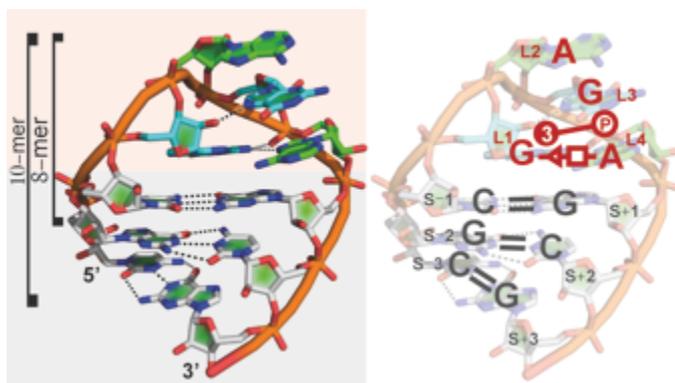

Figure 1. 3D (left) and secondary structures (right) of the studied GAGA hairpin, consisting of an A-RNA stem (grey labels and background) and tetraloop (red labels and background). Black dashed lines show the native hydrogen bonds and the signature interactions. The base-pairs and the base-phosphate interaction are annotated according to the standard nomenclature.[57, 63] We used two sequence lengths of the RNA hairpin: 8-mer TL (5'-gcGAGAgc-3') and 10-mer TL (5'-cgcGAGAgcg-3').

## SIMULATION SETTINGS

**T-REMD settings.** The T-REMD simulations were carried out using the AMBER suite of programs with several combinations of nucleic acid force fields, namely using ff99bsc0$\chi_{OL3}$ force field[32, 35, 38] (standard AMBER ff10-ff14 force field for RNA, for simplicity henceforth denoted as $\chi_{OL3}$), AMBER ff99 with the Chen&Garcia[33] modification (ff99Chen), and $\chi_{OL3}$ with Van der Walls modification of phosphate oxygen developed by Case et al. for phosphorylated aminoacids (labelled as $\chi_{OL3}$CP).[48] The Case phosphate parameters were implemented as introduced by Mlýnský et al.[34] including adjustment of torsion parameters affected by altered 1-4 Lennard-Jones interactions due to VdW modification, so that they resemble the original torsion profiles as closely as possible (AMBER library file of this force field version can be found in Supporting Information). We used 64 replicas in all T-REMD simulations. The temperatures of the 64 replicas spanned the range of 278-~500 K and were chosen to maintain an exchange rate of ~25%. The T-REMD simulations were performed at constant volume (using the NVT ensemble in each replica), with long-range electrostatics calculated using PME with 1 Å grid spacing and 10 Å real-space cutoff. Langevin dynamics with a friction coefficient of 2 ps$^{-1}$ was used as a thermostat in all replicas, and the exchanges were attempted every 10 ps. Each replica was simulated at least for 1 µs, while in cases, where we observed some folding events or in the simulation started from the folded state we extended the simulation time-scale to 2 µs per replica (see Table 1). The total simulation time accumulated by all T-REMD simulations was 896 µs.

**Table 1.** List of all simulations of 8-mer GAGA TL with sequence 5'-gcGAGAgc-3' and 10-mer GAGA TL with sequence 5'-cgcGAGAgcg-3'. $\chi_{OL3}$-neut.-fold and $\chi_{OL3}$-neut.MetaD simulations started from the folded state, while all others were initiated from the unfolded state.

| System | Method | Ionic conditions | Force field | Label | Length |
|---|---|---|---|---|---|
| 8-mer | T-REMD | 1M KCl | $\chi_{OL3}$ | $\chi_{OL3}$-1M | 1 µs x 64 |
| | | | $\chi_{OL3}$+HBfix 1 kcal | $\chi_{OL3}$HB-1M | 2 µs x 64 |
| | | | $\chi_{OL3}$+HBfix 0.5 kcal | $\chi_{OL3}$HB$_{0.5}$-1M | 1 µs x 64 |
| | | | ff99Chen | ff99Chen | 2 µs x 64 |
| | | | $\chi_{OL3}$-CP | $\chi_{OL3}$CP-1M | 2 µs x 64 |
| | | | $\chi_{OL3}$-CP+HBfix 1 kcal | $\chi_{OL3}$CP-HB-1M | 2 µs x 64 |
| | | K$^+$ net-neutral | $\chi_{OL3}$ | $\chi_{OL3}$-neut. | 1 µs x 64 |
| | | | $\chi_{OL3}$ | $\chi_{OL3}$-neut.-fold | 2 µs x 64 |
| | REST2 | 1M KCl | $\chi_{OL3}$ | $\chi_{OL3}$-1M-REST | 1 µs x 64 |
| 10-mer | T-REMD | K$^+$ net-neutral | $\chi_{OL3}$ | $\chi_{OL3}$-neut.-10mer | 1 µs x 64 |
| | WT-MetaD | Na$^+$ net-neutral | $\chi_{OL3}$ | $\chi_{OL3}$-neut.MetaD | 3 µs x 2 |

**REST2 settings.** The replica exchange solute tempering (REST2) simulation[11] was performed with 32 replicas at 300 K with the $\chi_{OL3}$ force field and scaling factor (λ) values ranging from 1 to 0.6057. Thus the effective solute temperatures (T/λ) ranged from 300 to ~500 K, i.e., in the range corresponding to our T-REMD simulations. The starting structures were identical to the above-mentioned unfolded states for T-REMD simulations (replicas 33 to 64, see Figures S3, S4, and S5 in Supporting Information). The values of λ were chosen to maintain an exchange rate ~25%. Exchanges were attempted every 10 ps.

**Well-tempered metadynamics settings.** The WT-MetaD simulations started from the native structure of 10-mer GAGA TL. At the beginning, the structure was equilibrated using a 5 ns NpT MD run with the Nosé-Hoover thermostat and isotropic Parrinello-Rahman (Andersen) barostat. After equilibration, two independent 3-μs-long WT-MetaD simulations were performed with a 2 fs time step at a temperature of 300 K using two CVs. The first CV involved all native hydrogen bonds (called $H_{core}$ CV) occurring in GAGA TL (both in stem and loop regions) and the second CV accounted for the all atom root-mean-square-deviation (RMSD) of the first ($G_{L1}$), third ($G_{L3}$), and fourth ($G_{L4}$) nucleotides of the loop from the native state. The WT-MetaD simulation with the Gromacs code 4.5.5[64] was performed using the PLUMED 1.3 plugin.[65] This simulation setup corresponded to the settings published in ref.[31] and in fact one of the present simulations was a prolongation of the simulation published there.

**HBfix: local potential energy function selectively supporting native hydrogen bonding.** To elucidate the influence of the strength of hydrogen bonds on folding of the GAGA TL, namely to test whether underestimated hydrogen bonding in the force field might affect propensity of the folding, we performed a set of T-REMD simulations with an additional locally-acting potential, increasing the stability of the base pairing and the signature interactions. The additional potential, henceforth denoted HBfix, is applied to heavy-atom distances of the *native* hydrogen bonds. Thus, six hydrogen bonds are involved in base pairing in the stem of 8-mer TL together with further three signature interactions of the loop (see Supporting Information for more details of the settings of the restraints in AMBER simulations and list of used native hydrogen bonds).

This short-range potential is composed of two flat-bottom harmonic restraints with linear extensions having opposite signs of their curvatures (Figure 2). The slopes of the linear extensions of the restraints are equal except for the sign and thus cancel each other. As a consequence, these restraints sum into a constant everywhere except for a small region, where at least one of the flat-bottom restraints is quadratic. In our particular case, the additional potential is non-constant from 3 to 4 Å, where it climbs from 0 kcal/mol to η kcal/mol, where the η value is a controllable parameter stabilizing the given hydrogen bond. As noted above, the potential acts locally. Outside the interval 3 to 4 Å, it does not result in any forces and thus does not perturb the equilibrium distribution. It stabilizes the H-bonds once formed but does not promote the target interactions when the interacting atoms are not in a close contact. It results in a short-range native-interaction energy bias that aims to compensate under-stabilization of the H-bonding by the AMBER force field due to missing polarization effects.[66] We performed T-REMD simulations of 8-mer with $\chi_{OL3}$ force field using two different values of the η parameter: a) 0.5 kcal/mol per hydrogen bond and b) 1.0 kcal/mol per hydrogen bond. Note that the simulations with a value of 0.5 kcal/mol did not lead to any visible increase of population of the folded state (the results of $\chi_{OL3}HB_{0.5}$ simulation can be found in the Supporting Information). In addition, we used the same η = 1.0 kcal/mol potential with $\chi_{OL3}$-CP force field (see Table 1). Apart from these additional HBfix potentials, the setup of these two T-REMD simulations was identical to the above-described protocol of other T-REMD simulations.

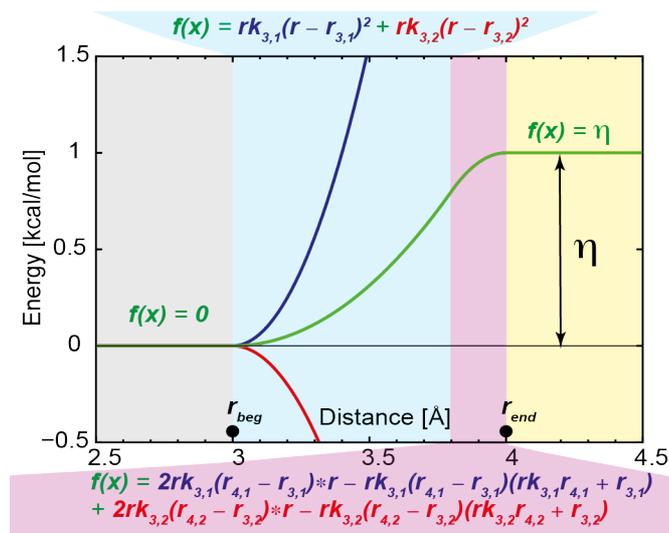

Figure 2. Description of the HBfix potential (green curve) applied to heavy-atom distances in native hydrogen bonds to strengthen these interactions. The potential is constant (i.e., it provides zero forces) at all distances except the narrow region between 3 and 4 Å. It is composed of combination of two flat-well restraints with opposite sign of curvature and linear extensions that cancel each other at distances above 4 Å (red and blue curves). The parameters $k_{3,1}$, $k_{3,2}$ and $r_{3,1}$, $r_{4,1}$, $r_{3,2}$, $r_{4,2}$ values stand for spring constants of the quadratic parts of the flat-well restraints and distances enclosing the quadratic region, respectively (see more details in Supporting Information).

**Data analysis.** The outputs of T-REMD and REST2 trajectories were sorted either to follow continuous trajectories or given temperatures, and Hamiltonians, respectively. All these trajectories were analysed with the Ptraj module of the AMBER package and the simulations were visualized using a molecular visualization program VMD.[67]

The estimations of error bars were calculated using bootstrapping.[68] The estimations of errors were obtained both by resampling time blocks in all 64 coordinate following replicas (identical time-blocks were resampled in all replicas simultaneously) as well as resampling this set of replicas used to finally obtain a resampled population at given temperature. In time domain we used a moving block bootstrap where optimal size of the blocks was selected to be close to $n^{1/3}$, where n is number of snapshots.[69] Subsequently, the replicas following coordinates were resampled (with replacement) and confidence interval with 5% level of significance was calculated from this resampled population estimated from 1000 resample trials. The resampling of time-blocks is typically used to estimate uncertainties originating from the limited sampling in time domain. However, this bootstrap implementation allows us in addition to distinguish between cases where the native state is formed only in one replica versus the native state being repeatedly formed in multiple replicas. Although in both cases one would observe both folded and unfolded structures in the reference replica (following the reference temperature), the latter scenario corresponds to populations that can be expected to be significantly closer to convergence. The above-described resampling of replicas takes this effect into account. This analysis is partly inspired by the analysis of continuous replicas that has been used in other works.[70]

For identification of the native state, we used two metrics: (i) presence of all native hydrogen bonds, i.e., all hydrogen bonds participating on the base pairing in the stem as well as signature interactions of the loop (presence of a hydrogen bond was inferred from a heavy-atom donor-acceptor distance less than 3.5 Å) together with all atom root mean square deviation (RMSD) calculated over all nucleotides except $A_{L2}$ ($A_{L2}$ stacking/unstacking fluctuations were observed in native state ensemble both theoretically and experimentally,[29, 32, 71-72] so its dynamics should not be considered as disruption of the native folded state) with cutoff 3 Å, and (ii) εRMSD metric. The latter metric was also used for clustering, see below. εRMSD is a recently introduced metric used to measure deviation between RNA structures.[73] This metric is based on selected geometrical properties and provides reliable information about differences in the base interaction network. The εRMSD was calculated using the baRNAba package [https://github.com/srnas/barnaba].

In order to identify dominant conformations sampled in the course of T-REMD simulations at a reference temperature of 300 K, we used a cluster analysis based on a clustering algorithm introduced by Rodriguez and Laio[74] in combination with the εRMSD metric. The main idea of this algorithm is that cluster centres are characterized by a higher density than their neighbours, and by a relatively large distance from any other points with higher local density. In order to apply the algorithm using the εRMSD metric, we had to introduce a slight modification of cluster selection, namely definition of cluster hull representing noise spread around given cluster (see Supporting Information for more details).

**RESULTS AND DISCUSSION**

Despite attempts to refine the RNA force fields (see the Introduction),[2-3, 32, 37-38] carefully performed MD studies continue to report persistent problems in description of many RNA molecules, including the smallest RNA systems such as tetranucleotides and tetraloops.[29, 52] Although recent reparameterizations are typically capable of correctly describing the native conformational basins as locally stable free energy minima (at least on the µs time-scale), robust sampling methods might indicate more complex imbalances in the force field if the native basin is not the global free energy minimum. The continuing problems in RNA simulations are not surprising considering the simplicity of the force field, large differences between force field and QM descriptions of nucleic acids,[55] and also the fact that essentially all refinements are still based on twenty-years-old second generation force fields.[2-3]

We aim to study the performance of selected current RNA force field versions for description of the conformational dynamics and folding of a small and structurally well-defined RNA motif, namely, the 5'-GAGA-3' TL. We assess the effects of the sequence length, ionic strength conditions, sampling technique, force field version, and definition of the native state. We want to separate problems caused by sampling from problems due to the force field, and identify how different force field terms affect the outcome of the simulations.

In total we analyzed eight T-REMD simulations with a cumulative time of 896 µs, one REST2 simulation with cumulative time of 32 µs, and two 3-µs-long WT-MetaD simulations, i.e., 934 µs of the simulation time in total (see Table 1).

**Definition of the native state may have a dramatic impact on the reported population of the folded state.** A key aspect in assessing the results is the definition of the native state. We strictly require that the TL must have not only the correctly paired WC stem, but also that all its signature interactions are properly structured. We classify the structures as misfolded TL when the stem is properly paired but the TL region is not properly structured. Note that some earlier studies considered the correct base pairing of the stem as sufficient to claim the TL folding while others counted even the spurious base zipper-like structures as folded.[75] When comparing with our stringent definition of native state, both later cases would be considered as "false-positives". Thus, definition of the folded state has decisive impact on the reported populations of folded structures. Similarly, as shown in the literature, RMSD can often be an unreliable criterion to separate folded, unfolded and misfolded states.[73]

We are aware of the fact that our definition of the folded state might be considered as excessively conservative. However, the requirement for the presence of all signature interactions in the folded state is motivated by strict conservation of the GNRA TL conformation including preservation of all signature interactions in the structural database revealed by X-ray crystallography and NMR spectroscopy.[76-78] It is well established that, in general, biochemically relevant structures of recurrent RNA motifs such as the GNRA TLs are strictly defined.[20, 30, 57, 79-81] Indirect but essentially indisputable indication of that is the perfect correspondence between the signature interactions and the consensus sequences, known also as the isostericity principle of RNA evolution.[57, 79] It proves that the evolutionary pressure aims to achieve precisely structured UNCG and GNRA TLs. Although the spectroscopic studies[71-72, 82-83] reported some inherent flexibility of the TL folded state, the majority of the alternative conformations suggested by these studies including, e.g., base-fraying of the terminal base-pair or unstacking of the L2 nucleotide from the tip of the TL, still maintain all the signature interactions. They thus fit to our definition of the folded state. The only conformation suggested by fluorescence studies that is not compatible with structural data is "5'-stacked" conformation having L2, L3, and L4 nucleotides stacked on the 5'-end of the stem.[71-72] However, such structure has never been reported by X-ray or NMR, and we also did not find any support for it in the present simulations. The absence of this conformation in the structural database cannot be explained by the limitations of X-ray crystallography and NMR spectroscopy due to high abundance of the GNRA TL in the database. In other words, the absence of 5'-stacked conformation cannot be explained by e.g. crystal packing, as the GNRA was resolved in numerous structural contexts and at least some of them are not affected by any crystal contacts. Therefore, we require in our definition of the folded state strict presence of the signature interactions suggested by available structural data. It is in line with current dominant view in RNA structural bioinformatics and structural biology.[80] As noted above, we admit that our definition of the folded state may be to certain extent considered too strict and specific for the MD technique. However, we can afford to use such definition due to the unlimited structural resolution of MD. Nevertheless, we admit that our definition of folded state is not necessarily fully compatible with the criteria (order parameters) utilized by diverse experiments. Due to the indirect nature of the order parameters (measured signals) used in some experiments, it would not be trivial to find an exact fully quantitative correspondence between the ensembles sampled in the simulations and in the spectroscopic measurements.

**Sequence length affects convergence of T-REMD simulations.** The length (number of nucleotides) of the studied RNA hairpin might have significant effects on the propensity for folding observed in enhanced sampling simulations. Longer sequences (hairpins with more base pairs in the stems) are expected to be thermodynamically more stable. Therefore, assuming converged simulations, we should obtain a higher population of the folded state for longer sequences at the reference temperature. On the other hand, the unfolded state of the hairpin with longer sequence has a larger available conformational space. Thus, its folding events would face a larger entropic barrier compared to hairpins with shorter sequences. An increase of the entropic barrier can particularly complicate the T-REMD method, since the method does not improve sampling over entropic barriers and folding events are expected to be less frequent. This leads to slower convergence of simulations with longer stems and thus possible misinterpretation of the comparison between two simulations considering different sequence lengths.

In order to test the effects of the sequence length, we first performed 1-μs-long T-REMD simulations using $\chi_{OL3}$ force field and net-neutral conditions for the 8-mer (5'-gcGAGAgc-3') and 10-mer (5'-cgcGAGAgcg-3') TL hairpin sequences. Alternative folding simulations of these sequences were recently reported in literature.[29, 33, 75] It is worth to note that the $\Delta G_{300K}$ of folding at 300 K estimated by Turner parameters[84] equals to -0.9 and -3.5 kcal/mol for the 8-mer and 10-mer, respectively. Such values would correspond to a rather minor difference in the expected population of the native state, namely 82.0% and 99.7% (see Table 2) in the case of the 8-mer and 10-mer, respectively. It is worth to note that the Turner parameters[84] were derived from the optical melting experiments relying on a folded state definition that is not straightforwardly related to the structural data. On the other hand, as discussed in the previous section, structural and bioinformatics data give a significant support for high stability of native TL conformation including all signature interactions, and thus it is likely that high populations estimated by Turner parameters correspond mostly to the native TL fold including signature interactions.

**Table 2.** Thermodynamic parameters of both studied TLs (8-mer and 10-mer) estimated by Turner et al. 2004 parameters[84] and the corresponding estimate of ΔG of folding and expected population of the native state at 300 K.

|  | 8-mer | 10-mer |
| --- | --- | --- |
| $\Delta G_{310K}$ [kcal/mol] | -0.1±0.2 | -2.5±0.2 |
| $\Delta H$ [kcal/mol] | -24.1±2.4 | -34.7±2.9 |
| $\Delta S$ [cal/mol/K] | -77.3±7.8 | -104.0±9.5 |
| $T_m$ [K] | 311.8 | 334.1 |
| $\Delta G_{300K}$ [kcal/mol] | -0.9±0.2 | -3.5±0.2 |
| $p_{native,300K}$ [%] | 82.0±4.4 | 99.7±0.1 |

We have obtained two basic results. First, we did not observe a single folding event in the two T-REMD simulations of the 8-mer and 10-mer (we analyzed the folding events in terms of our definition of the folded state, see above; see Figures S1 and S2 in Supporting Information). This observation contrasts with our previously reported 500-ns-long simulation of the same 10-mer sequence, where we observed a single folding event (in one replica).[29] The difference might be explained by the stochastic nature of the simulations, and confirms our original conclusions that the previously observed folding event was fortuitous. In other words, the native state is rarely sampled on the (sub)microsecond time scale of T-REMD simulations with the given simulation settings, regardless of the sequence length.

The second observation was a major difference between the propensities of the two hairpins to reach the misfolded state with the properly paired stem and unstructured loop. While the 8-mer simulation revealed a clear minor population of the native stem base pairing (16 - 44 % at 300 K after bootstrapping, see Figure 3), the 10-mer simulation sampled mostly unstructured single stranded states with a complete absence of the folded stem with concurrent formation of all three base pairs. Thus, the 8-mer sequence apparently showed a faster convergence and all subsequent T-REMD runs were carried out using the 8-mer hairpin. Poor sampling of stem formation was also reported by Cheatham and co-workers in their recent REMD study on the UNCG TL. They then decided to restrain the stem as they were primarily interested in the structure of the apical TL part of the hairpin.[52]

**The ionic conditions ($K^+$ net-neutral vs. 1M-KCl) may have only a subtle effect on convergence of the T-REMD simulations.** Sampling in folding simulations of RNA TLs may also be affected by ionic conditions. Due to finite size of the simulation box, the concentration of monovalent counter-ions is often above physiological concentration even in net-neutral conditions. Thus, net-neutralization should be sufficient to stabilize the negatively charged RNA backbone.[40, 44] On the other hand, the presence of divalent ions or very high concentration of monovalent ions might significantly affect both kinetics and thermodynamics of RNA folding.[85] Because of the known difficulties in modeling divalent ions with non-polarizable force fields,[86] we have limited our investigation to monovalent ions only.

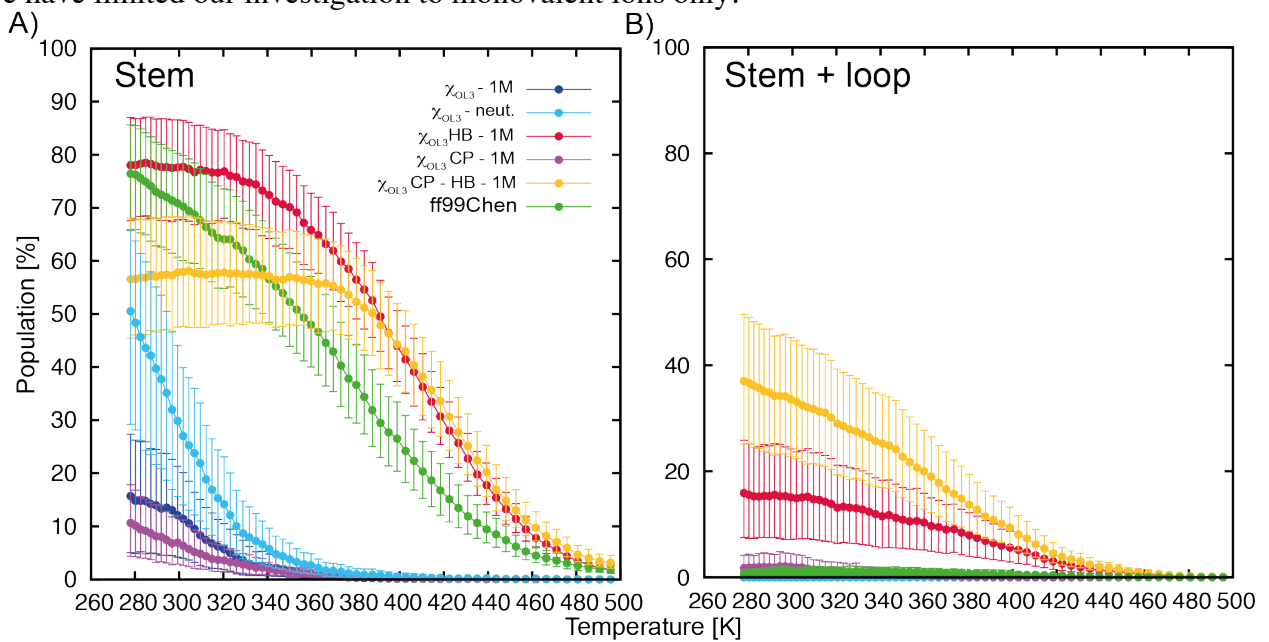

Figure 3. The population of A) misfolded structures with only folded stem (including the terminal pair) and B) native structures as a function of temperature calculated over all T-REMD 8-mer simulations. The error bars were calculated by bootstraping (using resampling of both time-blocks and coordinate following replicas, see Methods). Note that several simulations shown in the figure are discussed later in the text.

In order to examine the effect of ionic conditions, we compared T-REMD simulations of 8-mer GAGA TL in $K^+$ net-neutralizing condition (the simulation discussed in the previous paragraph) and 1M KCl excess salt. Similarly to the $K^+$ net-neutral simulation, we did not observe any complete folding event even in the 1M KCl salt excess, thus the native state is not sampled (or might be sampled only rarely) in T-REMD simulations on the microsecond time-scale regardless of the ionic conditions. Nonetheless, the simulations revealed slightly different propensity of the stem population, however, bootstrapping analyzes showed that the statistical significance of this difference was disputable. In particular, the stem was populated to 16-44% in net-neutral condition, while a slightly diminished population of 4-21% was observed in 1M KCl salt excess at 300 K (see Figure 3).

Detailed analysis revealed that the population of paired stems originated dominantly from preservation of prefolded stems structures present in some replicas in our starting pool of conformations. In other words, we faced some starting structure bias and limited convergence of some replicas (see Figure S2 in Supporting Information). As explained in the Methods, the starting structures were taken in 10 ns intervals from a standard 640-ns-long MD simulation at 300 K. Although this simulation started from a straight single-stranded conformation, spontaneous temporary ~120-ns-long formation of the stem appeared along this trajectory and, consequently, 13 replicas (~20%) started from such misfolded conformations, which were annotated by clustering analyzes as four-purine stack (see the corresponding section bellow). Some of these replicas lose the paired stem rapidly, while some of them were trapped in it for the entire simulation likely because they occupied mostly temperatures below 300 K, and their unfolding was thus slowed-down by low temperature.

The time-scales required for relaxation of this starting structure bias differed between the studied ionic conditions (see Figure S2 in Supporting Information). While 1M KCl salt excess simulations escaped from the four-purine stack misfolded state on tens-of-ns time-scale in all replicas, most of the replicas of the $K^+$ net-neutral simulation required hundreds of nanoseconds to relax this starting structure bias and some of them even remained trapped in this state. Therefore, the high population of the folded stem mostly originated from the starting structure bias. The observed difference between net-neutral and 1M KCl ionic conditions can be dominantly attributed to the different life-time of the four-purine stack misfolded state in different ionic conditions. It illustrates the complexity of sampling in T-REMD simulations which require in-depth monitoring of the individual replicas; for further details see Figure S2 in Supporting Information which gives time courses of the stem and stem+loop populations in the replicas. Nevertheless, it should be emphasized that some replicas formed the stem (and in particular the four-purine stack structure) spontaneously from true essentially straight single-strand configuration, so in the T-REMD 8-mer runs we observed also unbiased events of stem zipping (see Figure S2 in Supporting Information). Thus, we captured spontaneous stem zipping – unzipping processes and the four-purine stack definitely belongs to the folding landscape, as described by the force field. Taking all the data together, when we disregard the uncertainty in the four-purine stack population, the ionic conditions seem to have only a minor effect on the outcome of the GAGA T-REMD simulations. Nevertheless, the higher degree of trapping of the four-purine stack in net-neutral T-REMD suggests that under this ionic condition, RNA might be somewhat frustrated by limited amount of counter-ions. Thus, we hypothesize that it might be less flexible due to trapping in local free-energy minima, i.e., more prone to get trapped in some structures. It is worth noting that this suggestions, indirectly derived from the T-REMD simulations, could be consistent with lower flexibility of RNAs under low ionic strength reported by Woodson et al.[87] However, we re-emphasize that our simulations are still not quantitatively converged and further simulations on much longer time scales would be needed to clarify and quantify the effect.

**The RNA force field inaccuracy is likely responsible for unsatisfactory description of the GAGA TL folding.** In the above-described T-REMD runs, we did not observe any single folding event yielding into formation of a complete folded GAGA TL, though we would like to reiterate that one such event has been captured in our earlier study.[29] When considering all currently available data[29, 52] we suggest that computer simulations of the GAGA TL folding are still far from providing data consistent with the experimental observations. Even the latest $\chi_{OL3}$ force field is not sufficient to describe the GAGA TL folding from the unfolded state with a satisfactory population of the fully folded state, i.e., both with completely formed stem and loop. As noted in the Introduction, Chen and Garcia recently proposed variant of the AMBER force field with reparameterized van der Waals terms.[33] Using T-REMD simulations of three 8-mer TLs in 1M KCl salt conditions, including a different variant of GNRA TL, they reported dozens of spontaneous folding events.[33] This observation sharply contrasts with the behavior of the RNA GAGA TL in AMBER $\chi_{OL3}$ force field observed here and earlier,[29] although it should be noted that we use slightly different sequence of GNRA TL and different metric for definition of the folded state. The data present in the first part of our paper indicate that the difference between our simulations and those by Chen&Garcia can be attributed neither to the length of the hairpin sequence nor the ionic conditions. Therefore, it seems that the reparameterization of the van der Waals parameters might be vital for improvement of the description of the RNA hairpin folding. However, the ff99Chen force field has not been tested on any other RNA systems and subsequently other groups have demonstrated that these parameters may cause also undesired side effects for some other RNA systems like tetranucleotides, kissing loop complex or A-RNA duplexes.[52-53]

To obtain further insight, we decided to compare the $\chi_{OL3}$ and ff99Chen force fields for the 8-mer GAGA TL in 1M-KCl salt excess, using an identical metric of folding. We did not see frequent re-folding events comparable to those reported by Chen and Garcia. We carefully checked the force field settings to rule out that this discrepancy might originate from an erroneous implementation of the force field parameters. It should be, however, noted that here we used slightly different sequence of the TL, namely 5'-GAGA-3' instead of 5'-GCAA-3' used in the original study. Thus, a direct comparison of the present simulations with the preceding work is still to some extent limited and further comparisons would be desirable. Nevertheless, as can be seen in Figure 3, the ff99Chen parameters significantly increased the stability of the stem, but not the stability of the complete stem-loop structure. We observed many events of RNA stem zipping, i.e., formation of GC base pairs, but typically without proper formation of the native conformation of the GAGA loop region (defined strictly by its signature interactions, see Methods). At 300 K, we observed 0.0-0.7% and 0% population of the fully folded native state with ff99Chen and $\chi_{OL3}$, respectively (see Figure 3B). The stem was populated to 55-76% and 4-21%, respectively (see Figure 3A), confirming that ff99Chen force field has much better propensity to fold the stem.

**The reported differences in the literature may be affected by different definitions of the folded state.** In order to explain the significant discrepancy between various T-REMD runs discussed above, we focused on the description of the folded state. In our study we defined native (folded) state by simultaneous fulfillment of several conditions: i) presence of all signature interactions in the loop region, ii) presence of all hydrogen bonds participating on base-pairing in the stem, and iii) all-atom RMSD calculated over all nucleotides except the $A_{L2}$ below 3 Å. We excluded $A_{L2}$ from the RMSD calculation as it does not participate in any signature interaction of the TL and its reversible (un)stacking is an inherent part of structural dynamics of the folded TL.[29, 32, 71-72, 82, 88] This allows us to use a rather strict cutoff of 3 Å for the rest of the structure. This cut-off is justified based on the observed fluctuations in T-REMD simulations initiated from the native state (in which TL remain folded), which fluctuate in the RMSD range 0.7 to 2.2 Å. In contrast, Chen and Garcia defined the native state by 4 Å cutoff of all heavy-atom RMSD from the native state. We thus reanalyzed our simulations using all-atom RMSD with cutoff 4 Å as the only criterion. We found that the population of the folded state (within this simplified RMSD definition) significantly increased, namely to 12% and 38% in $\chi_{OL3}$ and ff99Chen 1M KCl 8-mer simulations, respectively. We conclude that, although the ff99Chen reparametrization significantly improves the stacking interaction, and stabilizes the native form of the stem, the stability of the tetraloop might be spuriously sensitive to the metric used to define the native state. A similar observation has been made by Cheatham et al. for the UNCG TL.[2] In addition, the conclusions based on RMSD metric can be misleading when important parts of the structures are excluded from the overall RMSD calculation. For example, all our above-mentioned T-REMD simulations revealed significant population with RMSD around ~2 Å when the RMSD was calculated only over heavy atoms of the stem and backbone atoms of the loop regions, respectively (see Figure S3 in the Supporting Information).

We would like to emphasize that all of the above analyses can be directly applied only to the simulations present in our work, since we could not analyze the trajectories obtained by the other groups.

The inadequacy of the simple RMSD metric is also discussed in the recent literature.[2, 73] Bottaro et al. suggested an alternative metric denoted εRMSD that is based on mutual position of coarse-grained nucleobases. This metric is significantly more robust compared to RMSD. Notably, in the case of 10-mer TL, even the spurious ladder-like structure[32, 45] is ~3.8 Å in RMSD from the native conformation.[29] The εRMSD is able to clearly distinguish this spurious state as a non-native conformation. The ladder-like structure of the TL gives a large εRMSD of ~1.2 while structurally similar RNA conformations are typically characterized by εRMSD below 0.5-0.7. We thus analyzed our T-REMD simulations using the εRMSD metric. The populations of the native state revealed by εRMSD metric is similar with populations obtained by our definition of the native state using full set of native hydrogen bonds in combination with the RMSD criterion. However, analysis of T-REMD simulations using εRMSD metric is significantly simpler compared to the cumbersome definition of the native state by signature interactions. Therefore, we suggest that εRMSD metric is a promising tool for analysis of nucleic acid simulations combining a simplicity comparable to the RMSD metric with significantly improved accuracy of identification of the native state from the non-native ones.

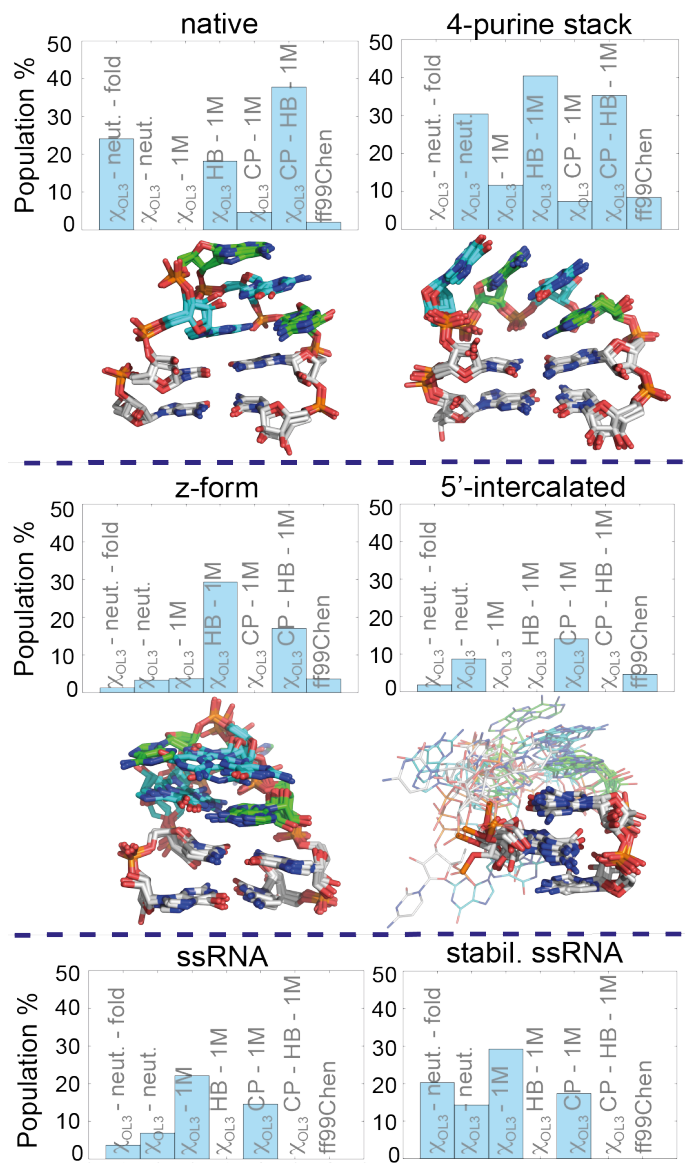

Figure 4. The most populated GAGA 8-mer TL clusters observed in T-REMD simulations at 300 K based on clustering using the εRMSD metric. The bars present for each particular cluster its populations in the indicated 8-mer T-REMD simulation (note that several simulations are discussed later in the text, see Table 1 for their overview). All graphs, except the two bottom-most ones which correspond to ssRNA clusters, are accompanied by the overlapped structures representing cluster centers obtained in all T-REMD simulations. Solid bonds (stick) depict those parts of the structures that are shared in the given cluster while variable parts are shown faded (lines).

**εRMSD-based clustering.** Motivated by the performance of the εRMSD metric in identification of the structural similarities, we performed a clustering analysis of all 300K replicas from the 8-mer T-REMD simulations using this metric.

We found that the hallmark of all T-REMD simulations starting from the unfolded state (i.e., all 8-mer T-REMD simulations, except the one starting from the folded state) was a significant population of the four-purine stack cluster. This was characterized by properly folded stem and continuous stacking of all four bases of the loop region as a 3'-overhang capping the stem (see Figure 4, which also includes simulations with force field modifications that are introduced later in the text). As detailed above, structures identified as four-purine stack cluster occured already in starting structures of 13 replicas, i.e., ~20% of total pool of the starting structures (see replicas 31-44 in Figure S4 in Supporting Information). This is because this state temporarily appeared during the MD simulation used for seeding the starting structures for the T-REMD simulations. Thus, high population of such structures during T-REMD runs could be affected by their presence in the initial ensemble (see Figure S4 in Supporting Information for time development of the clusters in coordinates following replicas). Nonetheless, although the populations might be affected by limited convergence and starting structure bias, we can still conclude that the four-purine stack is actually stabilized by the force field for two reasons. It was spontaneously formed in the classical simulation started from a straight single strand (used for seeding the starting conformations of the subsequent T-REMD simulations) and subsequently populated in T-REMD replicas. In addition, most importantly, we also observed several spontaneous folding events into such four-purine conformations during the T-REMD simulations from different conformations. In other words, we captured spontaneous unfolding and refolding of this particular conformation, i.e., there is an exchange with the unfolded ensemble in both directions. Thus, although the population of four-purine stack structure may be quantitatively overestimated due to their extensive presence in the set of starting replicas, the tendency of the force field to populate this conformation is indisputable.

Aside from the four-purine stack, almost all T-REMD simulations (except simulations stabilizing the native base pairs using the HBfix 1kcal/mol, introduced below) significantly populated the single stranded structures, i.e., structures lacking stem base pairing. Obviously, the ensemble of such structures is very diverse, so all clusters annotated as single stranded structures were sorted into two groups. In particular, single stranded structures stabilized by a wide variety of intramolecular interactions, mainly base-phosphate and sugar-phosphate (hydrogen bonding of 2'-OH with phosphate non-bridging oxygens) interactions were annotated as intermolecularly stabilized ssRNA (hencefort denoted as stabilized-ssRNA cluster, see Figure 4). The rest of single stranded structures, i.e., all structures having only stacking interactions between bases and lacking base pairing, base-phosphate and sugar-phosphate interactions, were further classified as the plain-ssRNA cluster (see Figure 4). The plain-ssRNA cluster includes not only A-form-like ssRNAs but also ssRNA structures with bulged bases. In addition, a prominent substate of the intramolecularly stabilized ssRNA state is 5'-intercalated cluster (see Figure 4), where the last two nucleotides at the 3'-terminus adopt a backbone conformation classified as 5d suite by Richardson nomenclature[89] while the 5'-terminal nucleobase is intercalated between them. It closely resembles a common simulation artifact observed for tetranucleotides.[52, 90-91] The last annotated misfolded cluster forms a stem, but in left-handed Z-form instead of A-form helix conformation and includes both stem guanines in *syn* orientation (Figure 4). Consistent with the other analyses, both $\chi_{OL3}$ simulations (net-neutral and 1M KCl salt excess) as well as the ff99Chen simulation reveal negligible population of the native folded state (below 1%, with zero population within error bars calculated by bootstraping).

Finally, we would like to note that we did not see any sign of ladder-like structure in any of our simulations, despite that all simulations were specifically monitored to detect such structures. This observation, together with several earlier-reported spontaneous reparations of ladders in $\chi_{OL3}$ simulations of short duplexes,[32] suggests that the formation of the ladder-like structures in $\chi_{OL3}$ force field documented by Cheatham et al. for UUCG TL at 277 K M-REMD reference replica might deserve a further analysis.[52]

**Several enhanced-sampling techniques (T-REMD, REST2 and WT-MetaD) point to the same results.** Due to the richness of the RNA conformational space, results of all simulation techniques, including enhanced sampling techniques, generally suffer from limited sampling. It seems to be a difficult task to achieve the ultimate convergence of the simulations even for small RNA systems.[49,52] Thus it is worthwhile to carefully analyze the convergence of the simulations and preferably use several independent techniques to obtain data that are as robust as possible.

The basic insight into estimation of accuracy achieved by the present level of sampling can be obtained by bootstrapping. Besides the standard block bootstrapping, where time-block data (at a given temperature) are resampled, we also resampled the set of coordinate-following replicas, from which the trajectory at given temperature is mixed (see Methods). Thus, the error estimations include not only inaccuracies originating from the limited time-scale, but also take into account inhomogeneities of the sampling among coordinate-following replicas. Note that this bootstrap analysis was able to reveal insignificance of the difference in folded stem population between 1M KCl salt-excess and $K^+$ net-neutral simulations that originated from the different level of relaxation of the starting structure bias (see above).

An ultimate requirement of the converged populations of the native state would be having identical results obtained by T-REMD simulations starting from fully unfolded and fully folded states. Thus, we performed another T-REMD simulation of GAGA TL using the $\chi_{OL3}$ force field and $K^+$ net-neutral condition starting from the folded state. The population of the native state at 300 K progressively decreased during the simulation and was never recovered after unfolding (see Figure S2 in Supporting Information). Therefore, we extended this simulation up to 2 µs per replica until the native structures were completely lost in all replicas and the population (at all temperatures/replicas) of the native state dropped to zero (note that Figure 4 reports total population calculated from the entire simulation time-scale). We cannot rule out that at longer time-scale we would be able to observe occasional short transient refolding events, as reported earlier.[29] However, the result is in clear qualitative agreement with the T-REMD simulations started from the unfolded state. All these T-REMD simulations together suggest that the canonical population of the native state under the employed force field description is negligible. In other words, the $\chi_{OL3}$ force field can reach and temporarily stabilize the native state of the GAGA TL but does not capture it as the most favorable conformational basin. In fact, our data suggest that the same applies for the ff99Chen force field. Thus, all currently available force fields fail to describe the correct global free energy minimum of this TL.

In order to obtain an alternative insight into the folding propensity of the GAGA TL in $\chi_{OL3}$ force field, we performed replica exchange solute tempering (REST2) simulation (see Methods). The REST2 run did not reveal any re-folding transitions (see Figure S5 in Supporting Information), supporting the conclusions based on the T-REMD simulations.

Finally, we also performed two independent well-tempered metadynamics simulations (WT-MetaD) of the 10-mer GAGA TL, where sampling of unfolding/re-folding dynamics was enhanced using two carefully chosen collective variables (CVs). The same CVs were used in our recent study[31] where we showed that these CVs are able to guide the simulation through reversible unfolding/re-folding events. Both 3-µs-long simulations resulted in qualitatively similar free energy landscape with clearly distinguishable free energy minima of single stranded, misfolded and native folded states (see Figure 5 and Figure S6 documenting structural dynamics in course of the WT-MetaD trajectories). Even our relatively long WT-metaD simulations are not able to provide rigorously converged estimates of the relative free energies of the different states. Nonetheless, they are in qualitative agreement with the T-REMD and REST2 simulations and suggest that the native folded state of the GAGA TL is not the global minimum under the force field description. The free energy of the native state was estimated by reweighting scheme to be ~3-4 kcal/mol above ssRNA and misfolded state (see Figure 5). It should be noted that while we speculated in our recent study[31] that the poor sampling of refolding events in WT-metaD simulations was caused either by still too simplified CVs or force field inaccuracies (or combination of both), the present data clearly suggest that the force field is not able to correctly describe GAGA TL folding.

Although all used enhanced sampling techniques, including simulations initiated from the folded state, point to the same conclusion, we would like to reiterate that none of the presented simulations is fully converged and our simulations could not reach steady state behavior.

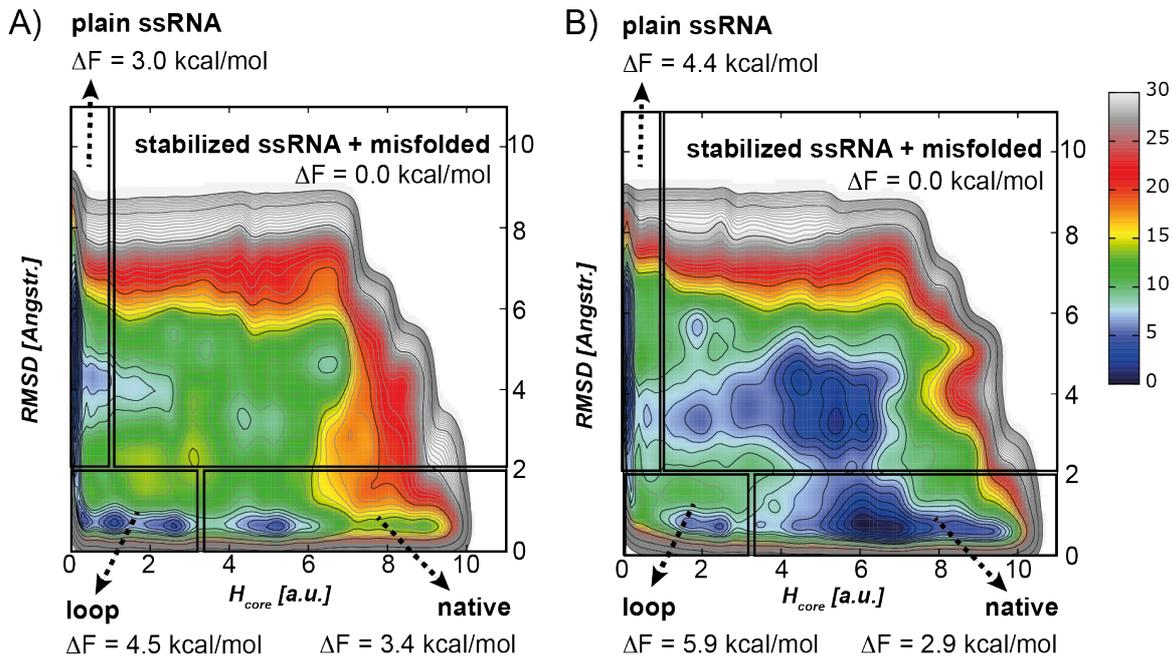

Figure 5. Free energy surface plots of two independent 3-μs-long WT-MetaD simulations of the GAGA TL using $H_{core}$ and RMSD of the TL as CVs. The FES is contoured by 1 kcal/mol (thin lines) and 10 kcal/mol (thick lines). Energies lower than 20 kcal/mol relevant for structural dynamics are colored. The regions depicted on the FES by black boxes correspond to the conformational states populated during the simulations. The free energy values of conformational states were estimated by reweighting scheme.

**Force fields may thermodynamically bias the simulations in favor of the ssRNA state by over-stabilization of BPh and sugar-phosphate interactions.** As mentioned above, the hallmark of the T-REMD simulations of GAGA TL is a significant population of the intramolecularly stabilized ssRNA state involving structures lacking base pairing but stabilized by intramolecular base-phosphate and sugar-phosphate interactions. Therefore, we tested the effect of the van der Waals reparameterization of the phosphate's oxygen developed by Case et al.[48] This reparameterization, albeit originally developed for phosphorylated aminoacids, was shown in our recent study to improve description of the structural dynamics of the hairpin

ribozyme active site.[34] Subsequently, Cheatham et al. showed that it also partially improved structural dynamics of tetranucleotides and UUCG TL.[52]

Here we used an implementation of these parameters described by Mlýnský et al.,[34] where VdW parameters of both bridging and non-bridging oxygens were modified, and all corresponding torsions were adjusted to take into account the effect of the modified 1-4 Lennard-Jones interactions (AMBER library file of this force field version can be found in Supporting Information). T-REMD 8-mer simulation using the combination of $\chi_{OL3}$ force field with the phosphate reparameterization was able to sample three folding events in a few replicas (see Figure S2 in Supporting Information), yielding a native state population comparable to the ff99Chen simulation (Figures 3 and 4). The initial success motivated us to extend this simulation up to 2 μs per replica. However, as in the ff99Chen simulation, bootstrap analysis showed that the population of the native state was still not significant (based on error bars representing the confidence interval calculated by bootstrapping at 5% level of significance, see Methods) at the present level of sampling; see error bars in the Figure 3 that reach zero population at all temperatures in both ff99Chen and $\chi_{OL3}$CP simulations (note that zero population was observed in case of all other above-mentioned $\chi_{OL3}$ simulations). This suggests that the phosphate reparameterization represents a step in the right direction but is not sufficient to radically change the force field performance. The results also suggest that excessive stabilization of stacking, which is corrected by the ff99Chen modifications, may only be part of the reason behind the poor sampling of the folded TL state with current force fields. Over-stabilization of the base-phosphate and sugar-phosphate interactions in the unfolded state may also be important.

**Local structure-specific biasing potentials might compensate for the effect of underestimated stability of native hydrogen bonds.** Another potential artifact of the empirical force fields that might preclude correct description of the RNA folding is underestimation of hydrogen bonding.[50, 92] In fact, part of the stem stabilization observed with the ff99Chen force field could be the increased stability of base pairing interactions, in addition to weakening of the stacking. Although the overstacking is a well-documented problem of contemporary force fields,[50, 92-93] a uniform rescaling of the VdW radii of the nucleobase atoms is inevitably accompanied by stabilization of base pairing, because the polar atoms involved in hydrogen bonding can come closer to each other, leading to a stronger electrostatic interaction. This is evident in the shorter heavy atom distances in hydrogen bonding interactions in the simulations with the ff99Chen force field (see Figure S7 in the Supporting Information), which might be an unintended consequence of the modification.

Considering the above results, we have attempted to stabilize the H-bonds with an additional term in the force field as a proof of principle. Instead of trying a global modification of the force field, which may be difficult to parameterize without incurring in undesirable side effects, we stabilized the native interactions locally and in a structure-specific manner. We aimed to strengthen hydrogen bonding interactions as gently as possible, to make minimal bias of the other parts of the folding landscape. We introduced a special additional HBfix potential affecting only a narrow range of heavy atom distances (between 3 and 4 Å, and supporting the hydrogen bonded state by simple potential energy bias as explained in Methods and Figure 2). Usage of 1 kcal/mol local bias per hydrogen bond applied to all native hydrogen bonds in the stem and loop region resulted in significantly increased population of the folded native state (Figures 3 and 4). Note, that in contrast to the other simulations, the error bars in this case did not reach zero and thus the observations are statistically significant over a wide range of temperatures (see

Supporting Information for alternative T-REMD simulation using bias of 0.5 kcal/mol per hydrogen bond). Note that a value of 1 kcal/mol per hydrogen bond supporting the base paring (3 kcal/mol for each GC base pair) approximately corresponds to the expected underestimation of the intrinsic hydrogen bonding interaction energy in the Cornell et al. based force fields, relative to quantum chemistry calculations.[50]

The slight improvement of the folding event sampling achieved with the Case et al.'s phosphate modification, and the significant effect of the HBfix potential motivated us to test these two modifications together. We hypothesize that while the phosphate's modification destabilizes unfolded state of GAGA TL, the HBfix stabilizes the folded state, so a cooperative effect might be expected. Indeed, we observed that the population of the folded native state in 2-μs-long $\chi_{OL3}$CP-HB simulation was significantly higher than the sum of the native state populations of $\chi_{OL3}$CP and $\chi_{OL3}$-HB simulations (see Figure 3). Thus we can conclude that these modifications act in a cooperative manner.

Structure-specific modifications targeting the native interactions are obviously not fully satisfactory. They require a priori knowledge of the native structure to be embedded into the definition of the potential energy function, i.e., they cannot be part of a transferable force field. Here we use them primarily to demonstrate the impact of an improved hydrogen bonding term on force field accuracy. On the other hand, when satisfactory reparametrization of the general force field is difficult, structure-specific modifications can be useful in studies of specific biomolecular systems.

We have also tried to estimate the effect of inclusion of non-native base-base hydrogen bonds via the same potential by reweighting. It does not alter the qualitative results, namely, after reweighting the native state would remain populated to 9.3% and 7.8% in $\chi_{OL3}$HB-1M and $\chi_{OL3}$CP-HB-1M simulations, respectively; cf. with the data in Figure 3. Our results thus show that introduction of a modest hydrogen bond correction significantly improves the population of the native state. This suggests that in future work it may be desirable to optimize a specific short-range hydrogen bond term, which would be mimicking the missing polarization effects accompanying hydrogen bonding. Such an optimization, which is beyond the scope of this work, could draw on existing quantum chemistry calculations, as well as experimental data for model systems in solution.

**CONCLUSIONS**

We performed a series of enhanced sampling simulations of the GAGA RNA TL testing several force fields, simulation settings and different kinds of enhanced sampling techniques in their performance of the description of the folding process.

Despite the microsecond time-scale, we found that none of the utilized enhanced sampling techniques (T-REMD, REST2, or WT-MetaD) was able to converge the sampling of the GAGA TL conformational ensemble and thus the reported simulations could not reach a steady state behavior. In addition, it should be noted that simulations using different force fields, involving a different sequence length, or having different ionic conditions might converge on a different time-scale, for example, due to their different propensity to form the RNA stem. However, different techniques and simulations settings used in our study pointed to the same qualitative conclusions, which allowed us to qualitatively estimate the stability of GAGA native folded state in different conditions (force field, sequence length and ionic conditions).

Our simulations confirmed the observation suggested in recent literature[29, 31, 33, 49, 52] about the inability of the $\chi_{OL3}$[35, 38, 42] force field to correctly describe balance between folded and unfolded states of the GAGA TL. The $\chi_{OL3}$ force field provides reasonable description of the folded state basin and consequently stable MD simulations of the TL on several-µs time scale as repeatedly shown in literature for many other RNA systems.[31, 41, 53, 94-95] However, the enhanced sampling techniques unambiguously reveal that the folded state of TL does not correspond to the global free energy minimum described by the force field. The simulations were able to form the correctly-paired A-RNA stem, but not the native TL structures. This means that we might face further problems in the force field, which do not appear to be correctable by refinements of the dihedral parameters. Thus, although the $\chi_{OL3}$ force field appears to do a very good job in simulations that start from known atomistic experimental structures of folded RNAs, it is not sufficiently robust to study RNA folding landscapes.

We have also tested the recently introduced modification of the van der Waals parameters by Chen&Garcia.[33] However, for the presently studied system, these modifications did not improve the folding landscape when considering the structure of the entire stem-loop.

When comparing different simulations we found that the common RMSD criteria of the folded state can be misleading and we strongly suggest monitoring the TL folding either with a full set of native interactions, or using the recently introduced εRMSD metric.[73] By specific examples we demonstrate how not-sufficiently-stringent definitions of the folded state may bias the interpretation of the folding simulations of RNAs.

One of the main aims of our work was to identify the possible sources of the force field artifacts that preclude correct description of the folded-unfolded balance of the TL. We show that at least two different effects likely contribute to incorrect folded/unfolded free energy balance of the TL, namely, excessive stabilization of the unfolded ssRNA state by intramolecular base-phosphate and sugar-phosphate interactions, and destabilization of the native folded state by underestimation of the native hydrogen bond interactions, including the stem base pairing. The former problem can be partially alleviated by alternative phosphate's oxygen van der Waals parameters developed by Case et al..[48] Although this Van der Waals term modification cannot be considered as a breakthrough in the RNA simulations, its positive effect on RNA simulations was shown by us and others also for other RNA systems.[34, 49] Until now it does not seem to be associated with any side effects. On the other hand, a generalized correction for underestimation of base-base hydrogen bonding (not associated with some side effects for other interactions) remains a challenge. We have nevertheless obtained substantially improved folding simulations of the GAGA TL using weak local structure-specific biases supporting the native interactions. In addition, we found that modification of hydrogen bonding acts cooperatively with the refined phosphate's parameters. We suggest that due to the persistent and principal limitations of the RNA force fields, such system-specific corrections represent a viable option to obtain stable native structures, with minimal perturbation to the transferable force field. In the long run, it would be desirable to develop an additional force field term for hydrogen bonding which captures this effect in a transferable manner.

**SUPPORTING INFORMATION**
Supporting Information includes detail description of used enhanced sampling techniques, setting of classical MD simulation used to seed the starting structures of REMD simulations, histograms of RMSD and εRMSD over all replicas, time evolution of the stem/loop populations in T-REMD and REST2 simulations, time evolution of structural dynamics in WT-MetaD simulations and detail description of the modifications applied to the clustering algorithm developed by A. Laio, detail description of the setting of the additional HBfix potential supporting hydrogen bonds, histograms of heavy atom distances of GC base pair hydrogen bonds in MD simulations compared to X-ray database, and AMBER library files including $\chi_{OL3}$CP force field.

**ACKNOWLEDGMENT**

This work was supported by grant P208/12/1878 (J.S., M.O., P.K.) from the Czech Science Foundation. Further institutional funding was provided by project LO1305 of the Ministry of Education, Youth and Sports of the Czech Republic (M.O., P.B., P.K.). JS acknowledges support by Praemium Academiae. R.B. was supported by the Intramural Research Program of the National Institute of Diabetes and Digestive and Kidney Diseases of the National Institutes of Health. G.B and S.B received funding from the European Research Council under the European Union's Seventh Framework Programme (FP/2007-2013) / ERC Grant Agreement n. 306662, S-RNA-S.